\documentclass[conference]{IEEEtran}
\IEEEoverridecommandlockouts
\usepackage{cite}
\usepackage{amsmath,amsfonts}
\usepackage{graphicx}

\usepackage{textcomp}
\usepackage{xcolor}
\usepackage{CJKutf8}
\usepackage{tabularray}
\usepackage{subfigure}
\usepackage{caption}
\usepackage{multirow}
\usepackage{bm}
\usepackage{amsmath,amsfonts}
\usepackage{algorithm}
\usepackage{algpseudocode}
\usepackage{array}
\usepackage[caption=false,font=normalsize,labelfont=sf,textfont=sf]{subfig}
\usepackage{textcomp}
\usepackage{stfloats}
\usepackage{url}
\usepackage{verbatim}
\usepackage{graphicx}
\usepackage{cite}
\usepackage{amsmath,amsfonts}
\usepackage{array}
\usepackage[caption=false,font=normalsize,labelfont=sf,textfont=sf]{subfig}
\usepackage{textcomp}
\usepackage{stfloats}
\usepackage{url}
\usepackage{verbatim}
\usepackage{graphicx}
\def\BibTeX{{\rm B\kern-.05em{\sc i\kern-.025em b}\kern-.08em
    T\kern-.1667em\lower.7ex\hbox{E}\kern-.125emX}}
\usepackage{balance}
\usepackage{amsmath}
\usepackage{makecell}
\usepackage{array}
\newcolumntype{C}{>{\centering\arraybackslash}X}
\usepackage{epstopdf}

\usepackage{changes}
\usepackage{textcomp}
\usepackage{xcolor}
\usepackage{tabularx,booktabs}
\usepackage[flushleft]{threeparttable}
\usepackage{booktabs,ragged2e}
\usepackage{multirow}
\usepackage{subcaption}
\usepackage{bm}

\usepackage{amsthm}

\allowdisplaybreaks[4]

\begin{document}

\setlength{\columnsep}{0.25in}

\begin{CJK}{UTF8}{gbsn}
\title{Evolving Token Communication with Parametric Memory Network}

\author{\IEEEauthorblockN{Weixuan Chen and Qianqian Yang\textsuperscript{\textsection}
}

\IEEEauthorblockA{{
}
{College of Information Science and Electronic Engineering, Zhejiang University, Hangzhou, China}\\
\{weixuanchen, \textsuperscript{\textsection}qianqianyang20\}@zju.edu.cn}

\thanks{This work is partly supported by the NSFC under grant No. 62293481, No. 62571487, No. 62201505, by the National Key R\&D Program of China under Grant 2024YFE0200802, and by the Zhejiang Provincial Natural Science Foundation of China under Grant No. LZ25F010001.  (Corresponding author: Qianqian Yang.)}

}

\maketitle

\begin{abstract}
Token communication has emerged as a promising framework for efficient wireless transmission by representing source data as compact semantic tokens. However, transmitting full semantic tokens still incurs considerable communication overhead. In this paper, we propose an evolving semantic token communication system with a parametric memory network over MIMO fading channels. Specifically, only an equal-length prefix of each semantic token is transmitted, which reduces transmission cost while preserving a consistent token structure for receiver-side recovery. At the receiver, a parametric memory network is introduced to reconstruct the missing suffix information from the received token prefixes, where semantic memory is stored implicitly in the network parameters. To realize this design, full semantic tokens are first organized into a codebook, and truncated tokens are paired with the codeword labels of their corresponding full tokens. Based on these token-label pairs, kNN-based teacher distributions are constructed to fine-tune a pretrained GPT-2-based recovery module, which learns to infer the codeword distribution of each incomplete token and recover the corresponding complete semantic token. In addition, an online evolution strategy is developed to periodically update the parametric memory network and the entire system using newly observed test samples, thereby improving adaptability under distribution shifts. Experimental results demonstrate that the proposed method consistently outperforms the existing evolving memory benchmark under different channel conditions and channel bandwidth ratios, with up to 1.09 dB PSNR improvement.
\end{abstract}


\section{Introduction}

Efficient transmission of high-dimensional source data over wireless channels is a fundamental challenge in modern communication systems. Different from conventional communication that emphasizes accurate bit-level delivery, semantic communication \cite{zhang2024semanticsurvey} aims to convey meaning-relevant information in a resource-efficient manner, and has shown great potential for multimedia transmission over bandwidth- and power-limited wireless channels. Recently, semantic token communication \cite{qiao2025token} has emerged as a promising framework, where source data are represented and transmitted as semantic tokens. By treating tokens as the basic communication units, it enables Transformer-based context modeling and provides a structured and flexible semantic representation for efficient compression and transmission. To fully exploit this advantage in wireless systems, it is crucial to reduce the transmission overhead of semantic tokens under limited communication resources.

To reduce such transmission overhead, existing works have mainly focused on adaptively controlling the extraction and transmission of semantic representations. In semantic communication,
Xie \textit{et al.} \cite{xie2023communication} developed a communication-efficient framework for distributed semantic transmission, in which a hierarchical vision Transformer and a task-adaptive translator are used to extract and transmit task-specific semantic representations in a coarse-to-fine manner. Rong \textit{et al.} \cite{rong2025semantic} introduced semantic entropy to quantify semantic importance, select informative semantic representations, and allocate them to higher-quality subcarriers for more efficient transmission. More recently, similar ideas have been extended to semantic token communication. Devoto \textit{et al.} \cite{devoto2026adaptivetoken} proposed an adaptive semantic token communication framework that dynamically discards less relevant tokens and adjusts the per-token compression ratio under varying resource and channel conditions. Ying \textit{et al.} \cite{ying2025joint} designed a joint semantic-channel coding and modulation scheme for semantic token communication, where tokens are mapped to modulated constellation points and the transmission process is further adapted by a rate allocator and a channel adapter. Despite these advances, existing methods mainly reduce transmission overhead through semantic representation selection, compression, and adaptive transmission, while the explicit recovery of missing semantic information from partially transmitted semantic representations remains largely underexplored.

Beyond adaptive transmission, another line of research enhances semantic communication with auxiliary memory or external knowledge. Xie \textit{et al.} \cite{xie2023memory} introduced a semantic communication framework with memory, where semantic representations from previous time slots are stored to support context-dependent tasks, and dynamic masking strategies are developed to improve transmission efficiency. Tang \textit{et al.} \cite{tang2024evolving} proposed an evolving semantic communication system with semantic caching, in which previously transmitted disentangled semantic representations are stored at both the transmitter and receiver, so that similar semantic representations can be replaced by indices in subsequent transmissions to continuously reduce communication overhead. From the perspective of knowledge-assisted decoding, Wang \textit{et al.} \cite{wang2023knowledge} designed a receiver-side knowledge-enhanced semantic communication framework, where a Transformer-based knowledge extractor retrieves relevant factual triples from noisy received representations and integrates them into semantic decoding to improve robustness. Zhou \textit{et al.} \cite{zhou2023cognitive} further developed cognitive semantic communication systems driven by knowledge graphs, where only key semantic triplets are transmitted and semantic errors are corrected through knowledge-graph inference and receiver-side semantic recovery. Such methods can improve communication efficiency and robustness, but they often introduce additional storage and retrieval overhead.

To address the above limitations, we propose an evolving semantic token communication framework with a parametric memory network for efficient wireless transmission over MIMO fading channels. The main contributions of this paper are as follows.


\begin{itemize}

\item We propose an evolving semantic token communication framework with parametric memory network.
Unlike existing methods that mainly reduce overhead by adaptive transmission or explicit memory assistance, the proposed framework performs prefix truncation at the transmitter and recovers missing semantic information at the receiver through a parametric memory network, thereby enabling efficient partial-token transmission.


\item We develop a parametric memory construction and learning strategy for semantic token recovery, inspired by the parametric memory paradigm in \cite{cao2025memory}. Specifically, a memory codebook is first constructed from full semantic tokens, while a datastore is built from truncated tokens to derive kNN-based teacher distributions that capture neighborhood semantic dependencies. Based on these components, a pretrained GPT-2-based recovery module is further fine-tuned to predict the corresponding codeword distributions from incomplete semantic tokens. In this way, semantic memory is distilled into the network parameters, enabling semantic token recovery without explicit online retrieval from external memory buffers or semantic caches.

\item We further design an online evolution strategy to improve adaptability under distribution shifts.
During testing, newly observed samples are periodically incorporated to update the parametric memory network and the entire system, so that semantic token recovery can be continuously refined as the data distribution evolves.

\item Experimental results demonstrate that the proposed framework consistently outperforms the existing evolving memory benchmark in \cite{tang2024evolving} under different channel signal-to-noise ratio (SNR) and channel bandwidth ratio (CBR) settings, achieving up to 1.09 dB PSNR improvement.

\end{itemize}

\section{System Model}

\begin{figure*}[t]
\begin{center}
\centerline{\includegraphics[width=0.99\linewidth]{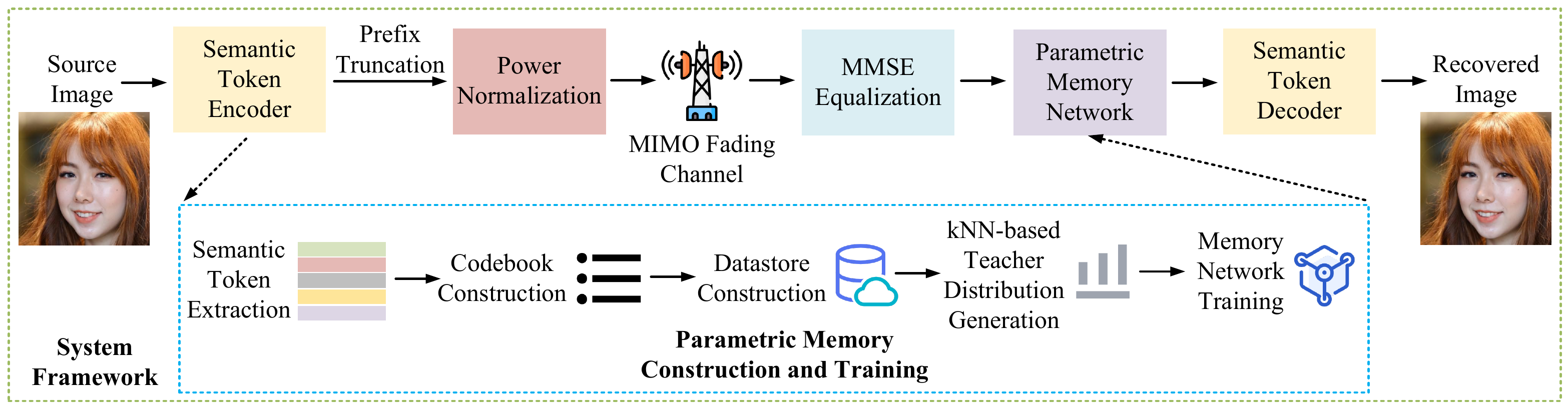}}
\caption{Overall framework of the proposed evolving semantic token communication system with a parametric memory network. The blue dashed box shows the construction process of the parametric memory network.}
\label{framework}
\end{center}
\vskip -0.2in
\end{figure*}

In this paper, we consider a semantic token communication system for image transmission over MIMO fading channels with a parametric memory network, as illustrated in Fig.~\ref{framework}. The system consists of a transmitter (Alice), a wireless channel, and a receiver (Bob). Specifically, the transmitter includes a semantic token encoder that transforms the source image into tokenized semantic information, while the receiver consists of a parametric memory network for semantic token recovery and a semantic token decoder for image reconstruction. The wireless channel is modeled as a MIMO Rayleigh fading channel with additive white Gaussian noise (AWGN). Given a source image $\mathbf{X} \in \mathbb{R}^{H \times W \times 3}$, Alice aims to convey its semantic information to Bob under limited transmission resources, so that Bob can reconstruct the source image $\hat{\mathbf{X}}$ from the transmitted semantic information.

Alice first encodes the source image into semantic tokens as
\begin{equation}
    \mathbf{Z} = f_{\mathrm{enc}}(\mathbf{X}),
\end{equation}
where $\mathbf{Z} \in \mathbb{R}^{L \times C}$ denotes the semantic tokens, with $C$ tokens each of length $L$, and $f_{\mathrm{enc}}(\cdot)$ denotes the semantic token encoder. To satisfy the transmission resource constraint, only a prefix of each token is preserved while the remaining suffix is truncated, yielding truncated semantic tokens $\mathbf{Z}_{\mathrm{tr}}$. This design is enabled by the receiver-side parametric memory network, which recovers the missing semantic information. The truncated semantic tokens are then normalized to satisfy an average transmit power constraint of 1 and converted into a complex-valued representation, resulting in the transmit signal $\tilde{\mathbf{Z}}$.

The transmit signal $\tilde{\mathbf{Z}}$ is then sent over a MIMO Rayleigh fading channel. We consider a MIMO system with $N_t$ transmit antennas and $N_r$ receive antennas, and adopt the conventional setting $N_t = N_r$. The received signal is given by
\begin{equation}
    \mathbf{Y} = \mathbf{H} \tilde{\mathbf{Z}} + \mathbf{N},
\end{equation}
where $\mathbf{H} \in \mathbb{C}^{N_r \times N_t}$ denotes the channel matrix, which represents the channel state information (CSI), whose entries follow independent and identically distributed complex Gaussian distributions $\mathcal{CN}(0, 1)$, and $\mathbf{N}$ denotes the AWGN, with each element following $\mathcal{CN}(0, \sigma^2)$. The channel signal-to-noise ratio (SNR) is defined as
$\mathrm{SNR} = 10 \log_{10} \left( \frac{P}{\sigma^2} \right) \; (\mathrm{dB})$,
where $P$ denotes the average transmit power constraint.

At the receiver, minimum mean-squared error (MMSE) equalization is applied to $\mathbf{Y}$ based on the estimated CSI $\hat{\mathbf{H}}$ to mitigate channel distortion. The equalized semantic information is given by
\begin{equation}
    \bar{\mathbf{Z}} = \hat{\mathbf{H}}^H \left( \hat{\mathbf{H}} \hat{\mathbf{H}}^H + \frac{\sigma^2}{P} \mathbf{I} \right)^{-1} \mathbf{Y},
\end{equation}
where $(\cdot)^H$ denotes the Hermitian transpose, $\mathbf{I}$ is the identity matrix, and $\sigma^2$ is the noise variance. The equalized semantic information $\bar{\mathbf{Z}}$ is then converted back to its real-valued representation, denoted by $\hat{\mathbf{Z}}_1$, for subsequent processing.

Since part of each semantic token is missing due to truncation, the equalized semantic information $\hat{\mathbf{Z}}_1$ is incomplete.
To address this issue, a parametric memory network is employed to recover the missing semantic information from the received token prefixes, which can be expressed as
\begin{equation}
    \hat{\mathbf{Z}}_2 = f_{\mathrm{mem}}(\hat{\mathbf{Z}}_1),
\end{equation}
where $f_{\mathrm{mem}}(\cdot)$ denotes the parametric memory network, and $\hat{\mathbf{Z}}_2$ denotes the recovered semantic tokens. The parametric memory network learns to infer the missing components of semantic tokens from their preserved prefixes, as detailed in the next section.

The recovered semantic tokens $\hat{\mathbf{Z}}_2$ are then fed into the semantic token decoder to reconstruct the image as
\begin{equation}
    \hat{\mathbf{X}} = f_{\mathrm{dec}}(\hat{\mathbf{Z}}_2),
\end{equation}
where $f_{\mathrm{dec}}(\cdot)$ denotes the semantic token decoder, and $\hat{\mathbf{X}}$ denotes the reconstructed image at the receiver.

Under prefix truncation, the effective channel bandwidth ratio (CBR) is determined by the actual number of transmitted token elements. Let $L_{\mathrm{p}}$ denote the preserved prefix length of each semantic token and $C$ denote the number of semantic tokens. Then, the effective CBR is defined as
\begin{equation}
    \mathrm{CBR} = \frac{L_{\mathrm{p}} C}{2HWC_{\mathrm{img}}},
\end{equation}
where $H$, $W$, and $C_{\mathrm{img}}$ denote the height, width, and number of channels of the source image, respectively. Since RGB images are considered in this paper, $C_{\mathrm{img}} = 3$. Here, the factor $2$ comes from converting real-valued token elements into complex-valued transmitted symbols.

The reconstruction performance is evaluated by the peak signal-to-noise ratio (PSNR), defined as
\begin{equation}
    \mathrm{PSNR} = 10 \log_{10} \left( \frac{255^2}{\mathrm{MSE}} \right),
\end{equation}
where MSE denotes the mean squared error between $\mathbf{X}$ and $\hat{\mathbf{X}}$.

\section{Proposed Method}

\subsection{Transmitter}

\subsubsection{Semantic Token Encoder}

At the transmitter, we employ a Swin Transformer-based semantic token encoder \cite{2025swinjscc} to extract compact semantic information from the source image. Given an input image $\mathbf{X}$, the encoder first applies patch embedding to convert the image into a sequence of patch-level tokens. These tokens are then processed by a hierarchical Swin Transformer backbone \cite{liu2021swin} with multiple stages, so that semantic information can be progressively extracted at different spatial scales.

Each stage contains a series of Swin Transformer blocks, where each block consists of layer normalization, window-based multi-head self-attention, residual connections, and a multilayer perceptron (MLP). Regular window attention and shifted-window attention are alternately adopted to enable both local modeling within each window and cross-window information interaction. Meanwhile, patch merging is introduced between adjacent stages to progressively aggregate semantic tokens, enlarge the receptive field, and reduce the spatial resolution.
After the hierarchical processing, layer normalization is applied, followed by a linear projection layer that maps the backbone outputs into semantic tokens for transmission, resulting in the semantic tokens $\mathbf{Z}$.

\subsubsection{Prefix Truncation}

After obtaining the semantic tokens $\mathbf{Z}$, we apply prefix truncation before transmission to reduce the transmission cost under limited transmission resources. The key idea is to preserve only a prefix of each semantic token while discarding the remaining suffix. This design is motivated by two considerations. First, transmitting full semantic tokens results in high communication overhead. Second, since the receiver is equipped with a parametric memory network that can recover missing semantic information from partial tokens, the transmitted content can be significantly reduced without severely degrading performance.

Specifically, for each semantic token of length $L$, only the first $L_{\mathrm{p}}$ elements are preserved, where $L_{\mathrm{p}} = \max(1, \lfloor \rho L \rfloor)$ and $\rho \in (0,1]$ denotes the prefix keep ratio. The same prefix ratio is applied to all tokens, ensuring a consistent truncation pattern across the entire semantic representation. The remaining suffix entries are not transmitted, resulting in truncated semantic tokens $\mathbf{Z}_{\mathrm{tr}}$. Since the truncation pattern is predefined and shared by both the transmitter and the receiver, the receiver can reconstruct the full token structure by zero-padding the missing entries without requiring additional side information.

\subsection{Receiver}

\subsubsection{Parametric Memory Network}

To reduce the transmission overhead under constrained communication resources, only truncated semantic tokens are transmitted, and a parametric memory network is introduced at the receiver to reconstruct complete semantic tokens from partially received ones. 
Unlike conventional memory-based approaches \cite{wang2023knowledge,tang2024evolving}, which perform explicit retrieval from a datastore or codebook during inference, the proposed parametric memory network stores semantic memory implicitly in its learnable parameters, following \cite{cao2025memory}. 
Accordingly, the recovery of missing semantic information is accomplished through a forward pass of the parametric memory network rather than an online nearest-neighbor search, thereby improving efficiency and reducing inference-time complexity.

Architecturally, the parametric memory network operates in a token-wise manner. 
Given an incomplete semantic token of length $L$, the memory network treats it as a one-dimensional token sequence and feeds it into a pretrained GPT-2-based token recovery module. 
Specifically, each scalar element of the incomplete semantic token is first projected into a latent embedding space through a linear embedding layer, and the resulting sequence is then processed by the first three Transformer blocks inherited from a pretrained GPT-2 model. 
After sequence modeling, the hidden representations are aggregated by a pooling operation, and a linear prediction head is applied to produce a probability distribution over the memory codebook. 
The recovered semantic token is then obtained as the expectation of the codebook entries weighted by the predicted probabilities, enabling the missing suffix to be inferred from the preserved prefix.

Next, we describe how the proposed parametric memory network is constructed, as summarized in Algorithm~\ref{alg:memory_construction}. The overall procedure consists of four stages: codebook construction, datastore construction, kNN-based teacher distribution generation, and parametric memory network training. Starting from the full semantic tokens extracted by the pretrained semantic token encoder, we first construct a codebook to represent the semantic token space. Then, truncated tokens and their corresponding full-token codeword labels are organized into a datastore, based on which a kNN-based teacher distribution is generated. Finally, the parametric memory network is trained to predict a codeword distribution for each incomplete semantic token, and the trained parametric memory network is integrated into the receiver for semantic token recovery.

\begin{algorithm}[t]
\caption{Memory Construction of the Parametric Memory Network}
\label{alg:memory_construction}
\textbf{Input}: Training image set $\mathcal{D}_{\mathrm{tr}}$, pretrained semantic token encoder $f_{\mathrm{enc}}$, prefix keep ratio $\rho$, codebook size $K$, and number of nearest neighbors $k$ \\
\textbf{Output}: Trained parametric memory network $f_{\mathrm{mem}}$ and memory codebook $\mathcal{C}$

\begin{algorithmic}[1]
\State Extract full semantic tokens from $\mathcal{D}_{\mathrm{tr}}$ using $f_{\mathrm{enc}}$;
\State Construct the memory codebook $\mathcal{C}$ from the extracted full semantic tokens;
\State Generate truncated semantic tokens by applying prefix truncation with keep ratio $\rho$;
\State Build a datastore by pairing each truncated token with the index of the nearest codeword to its corresponding full token in $\mathcal{C}$;
\State Build a kNN index over the truncated tokens in the datastore;
\For{each truncated semantic token in the datastore}
    \State Retrieve its top-$k$ nearest neighbors from the datastore;
    \State Aggregate the codeword labels of the retrieved neighbors into a soft teacher distribution;
\EndFor
\State Train $f_{\mathrm{mem}}$ to predict the teacher distribution from each truncated semantic token;
\State Deploy the trained $f_{\mathrm{mem}}$ together with the codebook $\mathcal{C}$ for receiver-side semantic token recovery.
\end{algorithmic}
\end{algorithm}

\subsubsection{Semantic Token Decoder}

At the receiver, a Swin Transformer-based semantic token decoder is employed to reconstruct the source image from the recovered semantic tokens. 
The decoder follows a hierarchical architecture similar to that of the encoder, while performing progressive upsampling to recover the spatial resolution.

Specifically, the input tokens are first projected into a higher-dimensional feature space and then processed by multiple stages of Swin Transformer blocks. 
Patch reverse merging is introduced between adjacent stages to progressively increase the token resolution. 
Finally, the output tokens are reshaped into the image space to obtain the reconstructed image $\hat{\mathbf{X}}$.

\subsection{Training and Online Evolution Strategy}

\subsubsection{Training}

The proposed system is trained in three stages. 
In the first stage, a baseline semantic token communication model is trained without prefix truncation and without the parametric memory network. 
The training objective is the image reconstruction loss
\begin{equation}
    \mathcal{L}_{\mathrm{rec}} = \|\mathbf{X} - \hat{\mathbf{X}}\|_2^2.
    \label{eq:lrec}
\end{equation}

In the second stage, the parametric memory network is trained independently using the kNN-based teacher distributions constructed from truncated semantic tokens and their corresponding full-token codeword labels, as summarized in Algorithm~\ref{alg:memory_construction}. 
Let $\mathbf{p}^{\mathrm{T}}$ denote the teacher distribution and $\mathbf{p}^{\mathrm{M}}$ denote the distribution predicted by the parametric memory network. 
The corresponding training loss is defined as
\begin{equation}
    \mathcal{L}_{\mathrm{mem}} = \alpha \, \mathcal{L}_{\mathrm{KL}}(\mathbf{p}^{\mathrm{T}} \,\|\, \mathbf{p}^{\mathrm{M}})
    + (1-\alpha)\,\mathcal{L}_{\mathrm{CE}},
\end{equation}
where $\mathcal{L}_{\mathrm{KL}}$ denotes the Kullback-Leibler divergence, $\mathcal{L}_{\mathrm{CE}}$ denotes the cross-entropy loss with respect to the target codeword label, and $\alpha \in [0,1]$ is a balancing coefficient.

In the third stage, the trained parametric memory network is integrated into the first-stage pretrained system, and the entire framework is further fine-tuned under the prefix truncation setting using the reconstruction loss in \eqref{eq:lrec}.

\subsubsection{Evolution}

To improve adaptability under distribution shifts, we introduce an online evolution strategy for the proposed system. 
During testing, the system is updated after every 20\% of the test set has been processed. 
Specifically, after each such interval, the newly observed samples are combined with the original training set, and the memory construction procedure in Algorithm~\ref{alg:memory_construction} is re-executed to reconstruct the codebook, regenerate the teacher distributions, and retrain the parametric memory network. 
Subsequently, the third training stage is repeated on the updated dataset, where the retrained parametric memory network replaces the previous one, and the entire framework is further fine-tuned under the prefix truncation setting using the previous encoder and decoder parameters as initialization. 
This enables the proposed system to continuously adapt to the evolving data distribution and maintain effective semantic token recovery.

\section{Performance Evaluation}

\subsection{Experimental Settings}

\subsubsection{Datasets}

We conduct experiments on the FFHQ64 dataset, which is a $64 \times 64$ version of the FFHQ dataset \cite{karras2021ffhq}. 
The dataset is split into 60{,}000 training images and 10{,}000 test images.

\subsubsection{Training Settings}

The experiments are conducted under fixed channel settings with $\mathrm{SNR} \in \{0,5,10,15,20\}$ dB. 
The baseline channel bandwidth ratio is set to $\mathrm{CBR}=1/8$, and prefix keep ratios of $20\%$ and $40\%$ are adopted, corresponding to effective CBRs of $1/40$ and $1/20$, respectively. 
A $2 \times 2$ MIMO system is considered in the experiments. 
All experiments are conducted on an NVIDIA RTX A6000 GPU.

For the baseline model in the first stage, we adopt the base SwinJSCC architecture with three downsampling operations and a patch size of 2. 
The encoder adopts a three-stage hierarchical design with embedding dimensions of 128, 192, and 256, using 2, 2, and 6 Swin Transformer blocks and 4, 6, and 8 attention heads in the three stages, respectively, while the decoder follows a mirrored structure with embedding dimensions of 256, 192, and 128, using 6, 2, and 2 Swin Transformer blocks and 8, 6, and 4 attention heads, respectively. 
For both the encoder and decoder, a window size of 4 and an MLP expansion ratio of 4 are adopted, with QKV bias and patch normalization enabled. 
The baseline model is trained for 500 epochs using Adam with a learning rate of $10^{-4}$ and a batch size of 32.

For memory construction and parametric memory network training in the second stage, the memory codebook size is set to $K=10{,}240$. For kNN-based teacher generation, a FAISS index with IVFPQ is adopted, with 2048 centroids, a code size of 32, and 32 probes. The teacher distribution is constructed from the top-512 nearest neighbors with a temperature of 8.0, excluding the nearest sample itself. The parametric memory network is trained for 100 epochs with a batch size of 1024 and a learning rate of $10^{-4}$, and the balancing coefficient is set to $\alpha=0.5$.

In the third stage, the trained parametric memory network is integrated into the receiver, and the entire system is further fine-tuned under the prefix truncation setting. The optimization settings follow those in the first stage, except that the learning rate is reduced to $5 \times 10^{-5}$. The same parameter settings are used during online evolution.

\subsubsection{Benchmarks}

We consider three benchmarks for comparison.
The first benchmark, denoted as \emph{Swin Token Comm}, is a Swin Transformer-based semantic token communication system \cite{2025swinjscc} without prefix truncation and without the parametric memory network.
The second benchmark, denoted as \emph{Swin Token Comm with Prefix}, applies the same prefix truncation strategy to the Swin Transformer-based semantic token communication system, but does not employ a semantic recovery module at the receiver. 
The third benchmark, denoted as \emph{Evolving Model with Cache}, is adapted from \cite{tang2024evolving}, where a dynamic semantic cache is employed to reduce transmission overhead and is progressively updated during testing. For a fair comparison, we implement its evolving cache mechanism within the same Swin Transformer-based semantic token communication framework as our method. In addition, the entire benchmark system is allowed to evolve during the testing stage, consistent with our setting, such that the performance gap mainly reflects the difference in semantic memory and recovery mechanisms.

\subsection{Performance Comparison of Different Approaches}





\begin{figure}[t]
\begin{center}
\centerline{\includegraphics[width=0.98\linewidth]{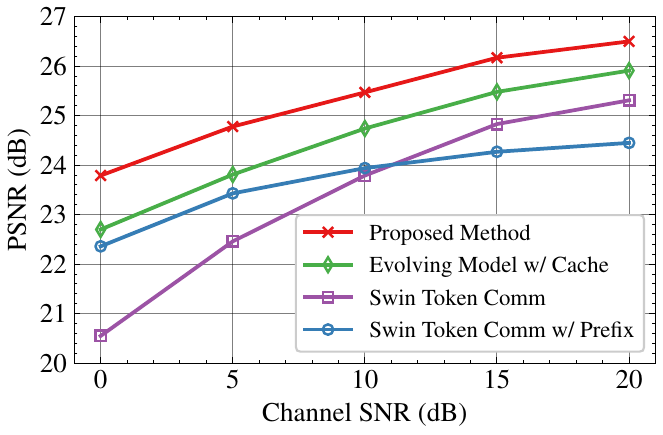}}
\caption{PSNR comparison of different approaches under different channel SNRs with $\mathrm{CBR}=1/40$, where $N_t=N_r=2$.}
\label{plot1}
\end{center}
\vskip -0.3in
\end{figure}

\begin{figure}[t]
\begin{center}
\centerline{\includegraphics[width=0.98\linewidth]{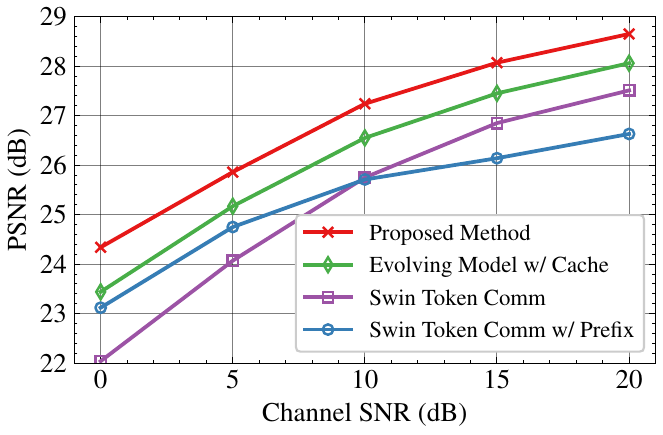}}
\caption{PSNR comparison of different approaches under different channel SNRs with $\mathrm{CBR}=1/20$, where $N_t=N_r=2$.}
\label{plot2}
\end{center}
\vskip -0.3in
\end{figure}

Figs.~\ref{plot1} and \ref{plot2} show the PSNR performance of different approaches under different channel SNRs with $\mathrm{CBR}=1/40$ and $\mathrm{CBR}=1/20$, respectively. For both CBR settings, the PSNR of all methods increases with the channel SNR, while the proposed method consistently achieves the highest PSNR over the entire channel SNR range.

For $\mathrm{CBR}=1/40$ in Fig.~\ref{plot1}, the proposed method achieves PSNR values ranging from 23.79 dB to 26.50 dB as the channel SNR increases from 0 dB to 20 dB. Its PSNR improvement over \emph{Swin Token Comm} ranges from 1.19 dB to 3.24 dB, showing a clear advantage under tight transmission constraints. Although \emph{Swin Token Comm with Prefix} improves over the baseline at low channel SNRs, its performance is still limited because it lacks an effective semantic recovery mechanism. \emph{Evolving Model with Cache} further improves the reconstruction quality, but it still performs worse than the proposed method across all channel SNR values.

For $\mathrm{CBR}=1/20$ in Fig.~\ref{plot2}, a similar trend is observed. The proposed method again achieves the best performance, with PSNR improving from 24.34 dB at 0 dB to 28.65 dB at 20 dB. Compared with \emph{Swin Token Comm}, the PSNR improvement ranges from 0.94 dB to 2.30 dB, while the improvement over \emph{Evolving Model with Cache} ranges from 0.59 dB to 0.90 dB. This shows that the proposed method maintains a stable advantage under different transmission conditions.

Comparing Figs.~\ref{plot1} and \ref{plot2}, all methods achieve higher PSNR at $\mathrm{CBR}=1/20$ than at $\mathrm{CBR}=1/40$, since more semantic information is retained during transmission. Overall, the proposed method consistently delivers the best reconstruction performance under both CBR settings, confirming the effectiveness of the proposed parametric memory network-based semantic recovery mechanism.

\section{Conclusion}

In this paper, we proposed an evolving semantic token communication system with a parametric memory network for wireless image transmission over MIMO fading channels. The proposed system reduced communication overhead by transmitting only an equal-length prefix of each semantic token, while reconstructing the missing suffix information at the receiver through a parametric memory network, in which semantic memory is implicitly stored in the learnable parameters. To realize this design, full semantic tokens were first quantized into a codebook, and each truncated token was associated with the codeword label of its corresponding full token. Based on these token-label pairs, kNN-based teacher distributions were constructed to fine-tune a pretrained GPT-2-based recovery module, enabling it to infer the codeword distribution of an incomplete token and recover the corresponding full semantic token. In addition, an online evolution strategy was introduced to periodically update the parametric memory network and the entire system using newly observed samples, thereby improving adaptability under distribution shifts. Experimental results show that the proposed method consistently outperforms the existing evolving memory benchmark under different channel SNR and CBR settings, achieving up to 1.09 dB PSNR improvement.

\bibliographystyle{IEEEtran}

\bibliography{myref}

\vspace{12pt}
\end{CJK}
\end{document}